\documentclass[12pt]{JHEP3}
\usepackage{mathrsfs}
\usepackage{amsmath,amssymb}
\usepackage{epsfig}
\usepackage{relsize}
\usepackage{graphicx}
\usepackage[all]{xy}
\def\N{${\cal N}=4$ }

\def\sl(2){\alg{sl}(2)}

\def\det{\hbox{det}}
\def\be{\begin{equation}}
\def\ee{\end{equation}}

\newcommand{\bea}{\begin{eqnarray}}
\newcommand{\eea}{\end{eqnarray}}

\newcommand{\bei}{\begin{itemize}}
\newcommand{\eei}{\end{itemize}}

\def\a {\alpha}

\def\la{\label}

\def\ov{\over}

\newcommand{\alg}[1]{\mathfrak{#1}}
\newcommand{\su}{\alg{su}}

\newcommand{\psu}{\alg{psu}}

\newcommand{\AdS}{{\rm  AdS}_5\times {\rm S}^5}

\newcommand{\sfrac}[2]{{\textstyle\frac{#1}{#2}}}

\newcommand{\bem}{\left (\begin{matrix}}
\newcommand{\eem}{\end{matrix} \right )}

\def\hstar{\,\hat{\star}\,}

\def\KI{K^{\rm I}}
\def\KII{K^{\rm II}}
\def\KIIt{\widetilde{K}^{\rm II}}
\def\KIII{K^{\rm III}}
\def\spm{\sqrt{\frac{x^+_i}{x^-_i}}}
\def\smp{\sqrt{\frac{x^-_i}{x^+_i}}}

\def\pI{  \prod_{i=1}^{K^{\rm I  }}}
\def\pII{ \prod_{i=1}^{K^{\rm II }}}
\def\pIII{\prod_{i=1}^{K^{\rm III}}}

\def\pIIt{ \prod_{i=1}^{\widetilde{K}^{\rm II }}}

\def\IG{\frac{i}{g}}

\author{Gleb Arutyunov$^a$\footnote{Email: G.E.Arutyunov@uu.nl, frolovs@maths.tcd.ie} {}\footnote{Correspondent fellow at Steklov
Mathematical Institute, Moscow.}\, \ and\,   Sergey Frolov$^{b\,\dagger}$
 \\ $^{a}$ {\it Institute for Theoretical
Physics and Spinoza Institute,\\ ~~Utrecht University, 3508 TD
Utrecht, The Netherlands} \\ $^b$ {\it Hamilton Mathematics Institute and School of Mathematics, \\
~~Trinity College, Dublin 2, Ireland} }

\abstract{ We discuss various aspects of 
excited state TBA equations describing the energy spectrum of the $\AdS$ strings and, via the AdS/CFT correspondence, the spectrum of scaling dimensions of ${\cal N}=4$ SYM local operators. We observe that auxiliary roots which are used to partially enumerate solutions of the Bethe-Yang equations do not play any role in engineering excited state TBA equations via the contour deformation trick. We further  argue that the TBA equations are in fact written not for a particular string state  but for the whole superconformal multiplet, and, therefore, the $\psu(2,2|4)$ invariance is built in into the TBA construction. }

\title{Comments on the Mirror TBA}

\preprint{
          \smaller{\smaller{\smaller{ITP-UU-11-08}}}\\[-.5ex]
          \smaller{\smaller{\smaller{SPIN-11-06}}}\\[-.5ex]
          \smaller{\smaller{\smaller{TCDMATH 11-04}}}\\[-.5ex]
          \smaller{\smaller{\smaller{HMI-11-03}}}}
\begin{document}

\renewcommand{\thefootnote}{\arabic{footnote}}
\setcounter{footnote}{0}


\section{Introduction and Summary}

The mirror Thermodynamic Bethe Ansatz (TBA) approach is the only  tool currently available to determine exact energies of $\AdS$ string states and, thanks to the AdS/CFT conjecture \cite{M},  scaling dimensions of \N SYM local gauge-invariant composite operators. In essence, the TBA is a set of coupled non-linear integral equations 
for the so-called Y-functions, whose solutions are expected to yield the spectrum of the corresponding string/gauge theory.

\smallskip
Although the TBA approach\footnote{We will not describe it here, referring the interested reader to the original literature \cite{Zamolodchikov90,Kuniba1} and recent
reviews \cite{Kuniba2,Arutyunov:2009ga,Bajnok:2010ke}. } has been successfully used in the case of  two-dimensional relativistic integrable models for quite some time \cite{Zamolodchikov90}, its application 
to the  $\AdS$ superstring\footnote{There is currently much evidence in favor of integrability of the $\AdS$ superstring and \N SYM, see the recent collection of reviews \cite{Brev} and references therein.} is not straightforward and requires a careful thought.   Importantly, the string sigma model is not Lorentz invariant on the two-dimensional world-sheet and, therefore, under 
the double Wick rotation, which is in the heart of the TBA construction,  it transforms into another model, termed in \cite{AF07} a {\it mirror}. The ground state 
energy of the original string model is then related to the free energy (or Witten's index, depending on the boundary conditions for fermions) of its mirror. 
In turn, the free energy and the TBA equations for the ground state  is derivable from the so-called string hypothesis \cite{Takahashi72}, which for the  $\AdS$ mirror 
model has been formulated in \cite{AF09a}.  In this way  the ground state TBA equations were obtained \cite{AF09b}-\cite{AF09d},\footnote{The final missing piece of the TBA which was the mirror dressing phase was provided by \cite{AFdp}.} and in \cite{FS} it was shown that 
the corresponding solution correctly reproduces the vanishing energy of the ground state that corresponds to the protected half-BPS operator of the gauge theory.

\smallskip

The importance of the ground state TBA equations lies in the fact that they admit a generalization to excited states by means of a 
contour deformation trick which is similar to the analytic continuation procedure of \cite{DT96}, see also  \cite{BLZe}.  The contour deformation relies on an assumption that the set of TBA equations is universal for any state of the model; excited states TBA equations may differ from each other only by a choice of integration contours of convolution terms, and by  analytic properties of Y-functions which determine driving terms in the TBA equations once the integration contours are taken back to the real line of the mirror model. TBA equations for string excited states in the $\sl(2)$ sector have been studied along these lines in \cite{GKV09b}-\cite {Frolov:2010wt}. 
 In general, Y-functions have quite intricate analytic properties which are currently under investigation \cite{AFS09}, \cite{CFT10}-\cite{AFT11}.

\smallskip

In this paper we continue studies of the mirror TBA approach and make three new observations. 

\smallskip

The first observation is on the origin of the large $J$ asymptotic solution.  As was emphasized in \cite{AJK},
in the large $J$ or small $g$ limit\footnote{Here $J$ is a charge of a string state and $g$ is the string tension which is related to the 't Hooft coupling
$\lambda$ as 
$\lambda=4\pi^2 g^2$.
}, the leading exponential correction to energies of string states should be given  by a proper generalization of 
L\"uscher's formula \cite{Luscher85}.  Such a generalization to the case of non-Lorentz invariant string sigma model and to string states containing many particles 
was proposed in \cite{BJ08} and used there to compute the four-loop anomalous dimension of the Konishi operator\footnote{In the context of the string sigma model L\"uscher's approach received recently a considerable attention, see the review \cite{Janik:2010kd} and references therein. The four-loop result for the anomalous dimension of the Konishi operator agrees with the field-theoretic computation
\cite{Sieg,Vel}.}. 
The corresponding energy correction is 
given in terms of $Y_Q$-functions that are in turn expressed via transfer matrices $T_{Q,1}(v)$, see eq.(\ref{YQasympt}). These transfer matrices  are associated with 
a scattering matrix which scatters a mirror theory $Q$-particle bound state of rapidity $v$ with string theory fundamental particles.  In what follows we refer to eq.(\ref{YQasympt})
as to the Bajnok-Janik formula.
We further recall that L\"uscher's formulae provide an approximation to the exact TBA equations when $Y_Q$-functions are small. Indeed, recently a perfect agreement has been found between
 L\"uscher's formulae at five loops \cite{BJ09,LRV09} and the corresponding predictions of the mirror TBA \cite{AFS10,BH10a}.
 
 \smallskip

 In addition to the main $Y_Q$-functions, the TBA equations also involve $Y_{\pm}$-, $Y_{M|vw}$- and $Y_{M|w}$-functions, as implied by  the string hypothesis 
 \cite{AF09b}.  Thus, to know the whole asymptotic solution, one has to also find the asymptotic expressions for   $Y_{\pm}$, $Y_{M|vw}$ and $Y_{M|w}$.
In this paper we show that the corresponding expressions  immediately follow from the Bajnok-Janik formula and the Y-system.
 We recall that the Y-system  is a set of functional relations between Y-functions which is obtained from the ground state TBA equations by applying a certain projection operation \cite{ZY}.
 In the context of the string sigma model the corresponding Y-system and the asymptotic solution were
 conjectured in \cite{GKV09}. The emphasis of our consideration here is on the fact that the asymptotic solution is {\it derivable} in a straightforward manner  from the Y-system and 
 the Bajnok-Janik formula, see Figure 1. In particular, in the process of the derivation, the Bazhanov-Reshetikhin formula \cite{BR} and Hirota equations \cite{Hirota} 
 relating various transfer matrices emerge naturally.

\begin{figure}
\begin{center}
 \includegraphics[width=0.9\textwidth]{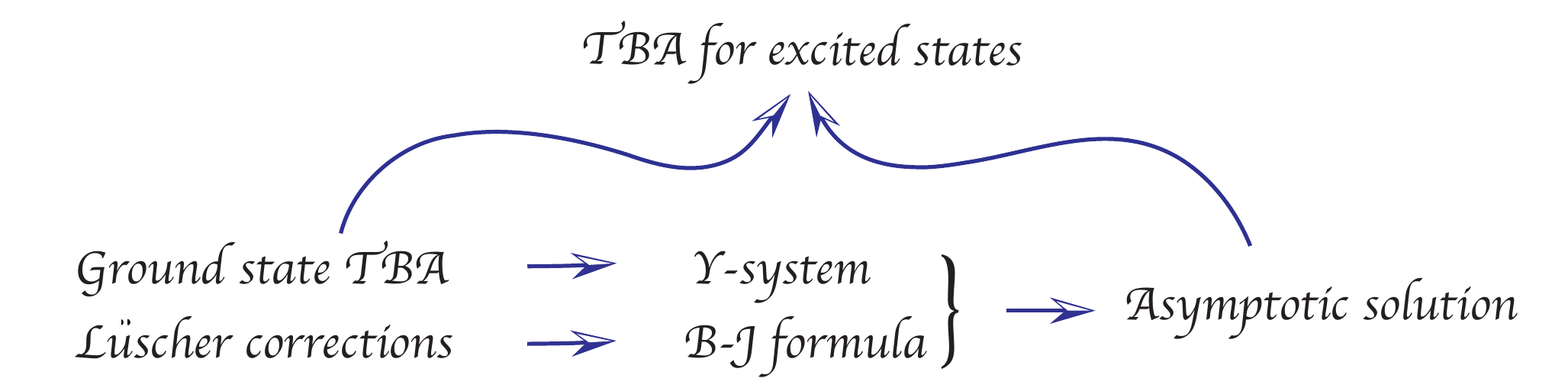}
\end{center}
\caption{The ground state TBA equations imply the Y-system. The Bajnok-Janik formula together with the Y-system 
leads to determination of the whole asymptotic solution. The ground state TBA equations together with the asymptotic solution allow one to engineer excited states TBA equations via the contour deformation trick.}
\label{fig:TBA-Y}
\end{figure} 

Our second observation concerns the construction of  excited TBA equations beyond the $\sl(2)$ sector. As was mentioned above, obtaining excited state TBA equations requires the detailed knowledge of the asymptotic solution. For generic states, the asymptotic solution is constructed in terms of transfer matrices that in addition to physical momenta 
$p_1,\ldots , p_N$ of string theory particles also involve auxiliary variables (roots) which satisfy the so-called auxiliary Bethe equations. We argue that for $J$ finite, a physical state 
is completely characterized by a set of charges it carries under the global symmetry group and by a set of momenta 
$p_1,\ldots , p_N$; auxiliary roots, as well as their Bethe equations, are invisible in the mirror TBA approach. Our arguing is based on the fact that all the transfer matrices and, therefore, the asymptotic Y-functions do not exhibit any singular behavior at locations of auxiliary Bethe roots. These roots satisfy auxiliary Bethe equations which guarantee regularity of transfer matrices.  As a matter of fact, it is the main rational behind the {\it analytic} Bethe ansatz \cite{Kuniba2} that Bethe equations are derivable from the requirement of analyticity of the corresponding transfer matrices. 
Of course, the presence of auxiliary Bethe roots affects analytic properties of  Y-functions, but only in an indirect way. More precisely, as a first step towards constructing excited states TBA equations, auxiliary roots corresponding to a given asymptotic state 
must be solved in terms of physical momenta and
substituted into asymptotic Y-functions, so the later  become functions of $p_1,\ldots, p_N$ and of rapidity $v$. Further, the physical momenta are determined from the main Bethe equations. 
Finally,  upon substituting  $p_1,\ldots, p_N$ into asymptotic Y-functions,
one has to determine their analytic properties as functions of $v$ and use them to engineer exact TBA equations via the contour deformation trick. We also point out, that the main (asymptotic)
Bethe equation for some $p_k$ implies the condition $Y_{1*}^o(p_k)=-1$, where  $Y_{1*}^o$ is an asymptotic $Y_1$-function analytically continued to the string theory region. For finite $J$
these asymptotic conditions are replaced by {\it exact} Bethe equations  $Y_{1*}(p_k)=-1$, where $Y_{1*}$ is an exact Y-function \cite{BLZe,DT96}. However, there are no any ``exact equations" for determining auxiliary 
Bethe roots, such equations are not there already for the asymptotic solution because of the above-mentioned analyticity of the asymptotic Y-functions.

\smallskip

Our third observation deals with the issue of the ${\rm PSU}(2,2|4)$ symmetry. This symmetry was an important ingredient for conjecturing the all-loop asymptotic Bethe ansatz \cite{BS,BES} based on the  $\psu(2|2)\oplus \psu(2|2)$-invariant S-matrix \cite{B05}. In the present paper 
we argue that ${\rm PSU}(2,2|4)$ is automatically built in into the mirror TBA approach, for a simple reason of being also the  symmetry of the 
asymptotic solution. It appears that asymptotic Y-functions are merely invariant under the action of supersymmetry generators, {\it i.e.},  they are the same for any member of a given superconformal multiplet.  Since the asymptotic solution is used to build exact TBA equations, the latter should be associated to the whole superconformal multiplet as well.  We also point out, that the relation $L_{TBA}=J+2$ between the TBA length parameter $L_{TBA}$ and the charge $J$
established in our previous work \cite{AFS09} has a interesting interpretation -- $L_{TBA}$ equals to the maximal $J$-charge in the supersymmetry multiplet under consideration.

\smallskip

The paper is organized as follows. In section 2 we give a derivation of the asymptotic solution from the Y-system and the Bajnok-Janik formula. 
We further explain how to engineer excited states TBA equations by means of the contour deformation trick. We also discuss the issue of auxiliary Bethe roots. 
In section 3 we discuss implementations of the ${\rm PSU}(2,2|4)$ symmetry in the mirror TBA approach.  In two appendices we collect the ground state TBA equations and discuss in detail duality transformation for transfer matrices, whose explicit expressions in various
gradings are also provided.

\section{TBA equations and asymptotic solution}

The mirror TBA approach is based on a bold assumption that  excited states TBA equations have the universal form  for any state. The TBA equations\footnote{For reader's convenience, the ground state TBA equations and following from them  Y-system equations are collected in appendix \ref{appYsys}.}  may differ from each other only by a choice of integration contours of convolution terms, and by  analytic properties of Y-functions which determine driving terms in the TBA equations once the integration contours are taken back to the real line of the mirror model. The driving terms are of the form $\pm\log S(u_*,v)$ where $u_*$ is either a zero, pole or $-1$ of a Y-function.  One can check that all the driving terms appearing via the procedure are annihilated  by the operator $s^{-1}$: $\log S\cdot s^{-1}(u_*,v)=0$ for $|v|\le2$, and, therefore, 
Y-functions which solve the excited states TBA equations also solve the associated Y-system equations.\footnote{The inverse is not true. Most solutions of the Y-system do not solve the TBA equations.}

\subsection*{ Asymptotic solution}

For large $J$ the energy of a string state can be found by solving the Bethe-Yang (BY) equations and taking into account the leading L\"uscher's exponential corrections. The BY equations depend on momenta of fundamental particles the string state is composed of (which could be complex if there are bound state particles), and on auxiliary roots. The auxiliary roots, however, are functions of the momenta, and, as a result, an $N$-particle string state is completely characterized by  the momenta $p_k$ or, equivalently, by the $u$-plane rapidity variable $u_k$, or by the $z$-torus variables $z_k$. Moreover, the imaginary parts of the momenta are determined by their real parts, and, therefore, the number of independent parameters is equal to the number of fundamental and bound state particles the string state is made of.  

\smallskip

Then, at large $J$ the $Y_Q$-functions become exponentially small 
and can be written in terms of the $\su(2|2)$ transfer matrices  defined in appendix \ref{appT} as follows \cite{BJ08}
\bea\la{YQasympt}
 Y_Q^o(v)&=&{\Upsilon}_Q(v)\,T_{Q,-1}(v)\,T_{Q,1}(v)\,, ~~~\\\nonumber
 {\Upsilon}_Q(v)&=&{e^{-J\widetilde{\cal E}_Q(v) }\ov  \prod_{i=1}^{N} S_{\sl(2)}^{1_*Q}(u_i,v)}=e^{-J\widetilde{\cal E}_Q(v)}\,  \prod_{i=1}^{N} S_{\sl(2)}^{Q1_*}(v,u_i)\,, ~~~\eea
where $v$ is the rapidity variable of the mirror $v$-plane and $\widetilde{\cal E}_Q$ is the energy of a mirror $Q$-particle. Here $S_{\sl(2)}^{1_*Q}$ denotes the $\sl(2)$-sector S-matrix with the first and second arguments in the string and mirror regions, respectively. The $\AdS$ string world-sheet S-matrix is a tensor product of two $\su(2|2)$-invariant ones, and this results in the appearance of two (in general different) transfer matrices 
$T_{Q,\pm1}(v)$ in \eqref{YQasympt}. The transfer matrices 
$T_{Q,\pm1}$ obviously depend on $u_k$, and the normalization of $T_{1,\pm1}$
is chosen so that if $u_k$ solve the BY equations then $ Y_{1_*}^o(u_k)=-1$.

\smallskip

The asymptotic $Y_Q$-functions \eqref{YQasympt} together with the Y-system equations completely fix the form of all the other Y-functions in the large $J$ limit. Indeed,  we first notice that for $|v|<2$ the prefactors ${\Upsilon}_Q$ in 
\eqref{YQasympt} satisfy the equations \cite{AF09d}
\bea\la{dL}
{{\Upsilon}_{Q}^{+}\, {\Upsilon}_{Q}^{-} = {\Upsilon}_{Q-1}{\Upsilon}_{Q+1}} \,.
\eea
Thus, for large $J$ the Y-system equation \eqref{YQ1} takes the form
\bea
{T_{1,-1}^+\,T_{1,-1}^-\ov T_{2,-1}}\,
{T_{1,1}^+\,T_{1,1}^-\ov T_{2,1}}
 &=& {\left(1 -  {1\ov Y_{-}^{(-)}} \right)
\left(1 -  {1\ov Y_{-}^{(+)}} \right) }\,,~~~~~~
\eea
and one finds the asymptotic solution for $Y_-$-functions
\bea\la{Ymasym}
Y_{-}^{(\pm)}=-{T_{2,\pm1}\ov T_{1,\pm2}}\,,\quad T_{1,\pm2}\equiv T_{1,\pm1}^+\,T_{1,\pm1}^--T_{2,\pm1}\,.
\eea
In the same way eq.\eqref{YQQ} allows one to get the asymptotic solution for $Y_{Q|vw}$-functions
\bea
Y_{Q|vw}^{(\pm)}&=&{T_{Q,\pm1}T_{Q+2,\pm1}\ov T_{Q+1,\pm2}}\,,\quad
T_{Q+1,\pm2}\equiv T_{Q+1,\pm1}^+\,T_{Q+1,\pm1}^- -T_{Q,\pm1}T_{Q+2,\pm1}\,.\la{Yvwasym}
\eea
Next, moving to eq.\eqref{Ym} for $Y_-$,  one determines the asymptotic 
$Y_{1|w}$-functions
\bea\la{Ywasym}
Y_{1|w}^{(\pm)}(v) &=& 
{T_{1,\pm2}^+T_{1,\pm2}^--T_{2,\pm2}\ov T_{2,\pm2}}={T_{1,\pm1}T_{1,\pm3}\ov T_{2,\pm2}} \,,\quad
T_{1,\pm3} \equiv  \left|
\begin{array}{ccc}
T_{1,\pm1}^{++}&1&0\\
T_{2,\pm1}^+&T_{1,\pm1}&1\\
T_{3,\pm1}&T_{2,\pm1}^-&T_{1,\pm1}^{--}
\end{array}
\right|.~~~~~~
\eea
Finally, eq.\eqref{Yvw1} for $Y_{1|vw}$-functions and 
eqs.\eqref{Yw1} and \eqref{YwM} for $Y_{Q|w}$-functions allow one to find
$Y_{+}$ and the remaining $Y_{Q|w}$-functions
\bea\la{Ypasym}
Y_{+}^{(\pm)} =-\frac{T_{2,\pm3}T_{2,\pm1}}{T_{3,\pm2}T_{1,\pm2}}\,,\quad Y_{Q|w}^{(\pm)} =\frac{T_{1,\pm Q\pm2}T_{1,\pm Q}}{T_{2,\pm Q\pm1}}\,,
\eea
where the transfer matrices $T_{a,\pm s}$ are expressed through $T_{a,\pm1}$ by means of the Bazhanov-Reshetikhin formula \cite{BR}
\bea\la{BRf}
T_{a,\pm s}(v)=\det_{1\leq i,j\leq
s}T_{a+i-j,\pm 1}\Big(v+\frac{i}{g}(s+1-i-j)\Big)\, . 
\eea
Here one assumes that $T_{a,\pm1}$ satisfy
the  conditions: $T_{0,\pm1} =1$ and $T_{a<0,\pm1}=0$.

\smallskip

It is worth mentioning  that by construction the asymptotic T-functions satisfy the T-system or Hirota equations \cite{Hirota}
\bea\la{Tasym}
T_{a,s}^+T_{a,s}^-=T_{a+1,s}T_{a-1,s}+T_{a,s+1}T_{a,s-1}\,,\quad a>0, s\neq 0\,,
\eea
where $T_{a,0}=1$. Clearly, the functions $T_{a,s}$ were constructed from $T_{a,1}$ in a purely algebraic manner, and it remains to be proven that they actually coincide with the transfer matrices corresponding to the representations $(a,s)$ of the centrally-extended $\psu(2|2)$. In the appendix we derive independently an explicit expression for $T_{1,s}$ which can be used to check numerically that it agrees with the one obtained from $T_{a,1}$ by means of \eqref{BRf}.

\smallskip

Obviously, in the large $J$ limit the negative $s$ T-functions are decoupled from the positive $s$ ones. For finite $J$ one takes the exact $Y_Q$-functions to be of the form  \cite{Kuniba1,Kuniba2}
\bea\la{YQexact}
 Y_Q&=&{\Upsilon}_Q\,{T_{Q,-1}\,T_{Q,1}\ov T_{Q-1,0}T_{Q+1,0}}\,, \eea
where $T_{Q,\pm1}$ reduce to the asymptotic ones and $T_{Q,0}$ reduce to 1 in the large $J$ limit. Then one  assumes that in addition to the Hirota equations \eqref{Tasym}   the T-functions satisfy the following equations\footnote{Since ${\Upsilon}_Q$ obey the discrete Laplace equations \eqref{dL}, they can be absorbed in the transfer matrices by a proper rescaling, see {\it e.g.} \cite{Ryo}.  }
\bea\la{Ts0}
T_{a,0}^+T_{a,0}^-=T_{a+1,0}T_{a-1,0}+{\Upsilon}_a\, T_{a,1}T_{a,-1}\,,\quad a>0\,,
\eea
and expresses all the other Y-functions in terms of $T_{a,\pm1}$ and $T_{a,0}$, or, equivalently, in terms of $T_{1,s}$. 
Introducing the functions
\bea\nonumber
&&Y_{Q,0}=Y_Q\,, \quad Y_{1,-1}=-\frac{1}{Y_-^{(-)}}\,, \quad Y_{1,1}=-\frac{1}{Y_-^{(+)}}\,,
\qquad Y_{2,-2}=-Y_+^{(-)}\, ,\quad Y_{2,2}=-Y_+^{(+)}\, ,\\
&&Y_{Q+1,-1}=\frac{1}{Y_{Q|vw}^{(-)}}\,,\quad
Y_{Q+1,1}=\frac{1}{Y_{Q|vw}^{(+)}}\, ,\qquad
Y_{1,-Q-1}=Y_{Q|w}^{(-)}\,,\quad
Y_{1,Q+1}=Y_{Q|w}^{(+)}\,,~~~~~
\eea 
and performing the rescaling mentioned in footnote 8, one can write the relations in the standard form  \cite{Kuniba1,Kuniba2}
\bea
Y_{a,s}={T_{a,s-1}T_{a,s+1}\ov T_{a-1,s}T_{a+1,s}}\,.
\eea
It is important to stress that the existence of T-functions $T_{a,\pm1}$ and $T_{a,0}$ which satisfy \eqref{Ts0} does not follow from anything we know about the $\AdS$ superstring or \N SYM. 
Fortunately, this assumption plays no role in constructing excited states TBA equations via the contour deformation trick.

\subsection*{ Engineering TBA equations  for any state}
In this subsection we formulate the general strategy for engineering excited state TBA equations via the contour deformation trick. 
It is basically the same as the one we used in \cite{AFS09} to analyze the states from the $\sl(2)$ sector. The main new ingredient  now is that the description of a generic state 
by means of the BY equations requires using not only momenta carrying Bethe roots but also a number of auxiliary ones. 

\begin{enumerate}
\item Start with the BY equations in any grading, {\it e.g.}, the $\sl(2)$ or $\su(2)$ ones, and choose a charge $J$,  a number of fundamental particles $N=\KI$ and a set of auxiliary roots numbers $\KII_\a$ and $\KIII_\a$ for the state/operator under consideration.  All the other 4 charges of the state are determined by $J$ and the auxiliary roots numbers. The canonical dimension of the primary operator or the energy of the string state at $g=0$ is given by 
$E_0=\Delta_0 = J+N$.

Solve the BY equations  and choose from many solutions the one which 
corresponds to the state of interest.
This state is characterized by a definite set of 
$g$-dependent momenta, and by the set of auxiliary roots numbers $\KII_\a$ and $\KIII_\a$.  Auxiliary roots are completely fixed by the momenta, and play no independent role in the description of the state.  
In practice a solution of the BY equations can be found only numerically for small $g$.

\item Compute asymptotic T- and Y-functions by using the set of momenta and auxiliary roots for the state, and find the location of  zeroes and poles of 
$1+Y_{a,s}$ and $Y_{a,s}$ functions in the mirror and string regions.  The asymptotic solution can be trusted for finite $J$ and small $g$.
Note that the grading of the transfer matrices should match the grading of the BY equations used.

\item Choose integration contours and 
 engineer TBA equations for the state so that the asymptotic TBA equations obtained by dropping the terms with $\log(1+Y_Q)$ are solved by the asymptotic solution for Y-functions.

\item The exact momenta of fundamental particles are found from the exact Bethe equations $Y_{1*}(p_k)=-1$ which are derived by analytically continuing the excited state TBA equation for $Y_1$. 
\end{enumerate}

Following the steps  one can write down (at least in principle) excited states TBA equations for an arbitrary string state or \N primary operator.

Some important comments are in order.

\bei
\item There are no equations to determine the exact location of the auxiliary roots. As we discuss in the next subsection, Y-functions are regular at the locations of auxiliary roots, and do not have neither zeroes, poles or $-1$'s there. 

\item Asymptotic Y-functions  have  zeroes, poles or $-1$'s at other locations, and exact Y-functions  have  similar properties too.
Exact locations of these zeroes, poles and $-1$'s are determined by analytically continuing the TBA equations for corresponding Y-functions to their (approximate) locations, and setting the Y-functions there to $0\,,\ \infty$ or  $-1$. 

\item  Some zeroes, poles or $-1$'s of asymptotic Y-functions might be spurious and be absent in exact Y-functions. 
For example, the analysis of a state from the $\su(2)$-sector composed of a fundamental particle and a two-particle bound state  shows \cite{AFT11}  that auxiliary Y-functions may have some zeroes, poles or -1's whose contributions to the TBA equations should not be included. These spurious zeroes, poles or $-1$'s are outside the physical strip and related to asymptotic $Y_\pm$ functions and probably to $T_{2,3}$ and $T_{3,2}$ which are on the boundary of the T-hook.

\item The energy spectrum computed by using the asymptotic 
Bethe equations is very degenerate. Most of this degeneracy is however lifted as soon as the leading exponential corrections are taken into account. 

As an example, let us consider the states which have $\KI=4$, $\KII_-+\KII_+=2$ and
$\KIII_\a=0$. There are obviously three possible $y$-root distributions: 
(i) $\KII_-=0, \KII_+=2$, (ii) $\KII_-=2, \KII_+=0$, (iii) $\KII_-=1, \KII_+=1$.
Since $\KIII_\a=0$  all the $y$-roots satisfy one and the same equation \eqref{BA1}. There are two solutions, $y_1$ and $y_2$, to this equation not equal to $0$ or $\infty$. Let one of the $y$-roots be equal to $y_1$, and the other to $y_2$. 
The next step is to find a solution to the main Bethe equation \eqref{BAmain}.  This equation involves all the $y$-roots, and, therefore, the solution is the same for any of the three states. Thus, the asymptotic energies of these states are the same.
The states (i) and (ii) correspond to conjugate representations 
and therefore have equal energies also for finite $J$ and $g$. On the other hand,  computing the $Y_Q$-functions for the third state one finds that they are different from those for the first two states. Thus, already the leading exponential correction would lift the asymptotic degeneracy of the spectrum. 

\eei  

We conclude this discussion with a word of caution. The procedure described above has  so far been tested only for states composed of fundamental particles with real momenta.  If some of the momenta are complex, {\it e.g.} there are bound states, 
then in the large $J$ limit some of the asymptotic $Y_Q$-functions constructed via eq.\eqref{YQasympt} develop singularities on the real line of the mirror theory. Therefore,   
strictly speaking, eq.\eqref{YQasympt} does not provide a large $J$ asymptotic solution of TBA equations.  
We believe however that for finite $J$ and small $g$ eq.\eqref{YQasympt} or its mild modification can still be used to construct TBA and exact Bethe equations.   
This issue is currently under investigation \cite{AFT11}.


 \subsection*{Auxiliary roots and T- and Y-functions}
 
 In this subsection we discuss whether the locations of auxiliary roots $y_k^{(\a)}$ and $w_k^{(\a)}$ are encoded in any way in analytic properties of  T- and Y-functions. A naive expectation would be that some of the auxiliary Y-functions should be equal to $-1$ just as $Y_{1_*}(u_k)=-1$ at the locations of the main Bethe roots. We will see that this is not the case and all T- and Y-functions are regular for any value of $v$ related to the auxiliary roots $y_k^{(\a)}\,, w_k^{(\a)}$ by the shifts of the form $in/g$ for integer $n$. Regularity of transfer matrices at locations of auxiliary Bethe roots is a well-known fact, and our discussion below is given mainly to illustrate our conclusions.
  
\medskip
 
 It is sufficient to consider only the roots with $\a=+$ which will be denoted as $y_k\,, w_k$.  For definiteness we discuss transfer matrices  in the mirror $v$-plane.  The same conclusions are reached by considering the analytic continuation of the transfer matrices to the string $u$-plane.

We begin the discussion with $T_{1,1}$. It is written in the form, see appendix \ref{appT}
\bea\la{t11}
T_{1,1}(v)={\mathscr N}_1(v)\Omega_1(v)\, ,
\eea
where the normalization factor 
\bea 
{\mathscr N}_1(v)=
\prod_{i=1}^{K^{\rm{II}}}{\textstyle{\frac{y_i-x^-}
{y_i-x^+}\sqrt{\frac{x^+}{x^-}}} }  
\eea 
is  necessary to provide a proper normalization of $Y_1$-functions,
and
\begin{eqnarray}\label{eqn;FullEignvalue2}
\Omega_1(v)=1 &+& \prod_{i=1}^{K^{\rm{II}}}{\textstyle
\frac{v-\nu_i+\frac{i}{g}}{v-\nu_i-\frac{i }{g}}}
\prod_{i=1}^{K^{\rm{I}}} {\textstyle{\frac{(x^--x^-_i)(1-x^-
x^+_i)}{(x^+-x^-_i)(1-x^+
x^+_i)}\frac{x^+}{x^-} }}-\\
& - & \prod_{i=1}^
{K^{\rm{I}}}{\textstyle{\frac{x^+-x^+_i}{x^+-x^-_i}
\sqrt{\frac{x^-_i}{x^+_i}} }}  \times\nonumber
\left\{\prod_{i=1}^{K^{\rm{III}}}{\textstyle{\frac{w_i-v
-\frac{2i}{g}}{w_i-v}}}+ \prod_{i=1}^{K^{\rm{II}}}{\textstyle
\frac{v-\nu_i+\frac{i}{g}}{v-\nu_i-\frac{i
}{g}}}\prod_{i=1}^{K^{\rm{III}}}{\textstyle{\frac{w_i-v
+\frac{2i}{g}}{w_i-v}}}\right\}\,  \nonumber
\end{eqnarray}
is normalized so that $\Omega_1(u_k)=1$. This guarantees that the asymptotic $Y_Q$-functions satisfy the conditions $Y_{1_*}(u_k)=-1$ if the main Bethe roots $u_k$ solve the BY equations.

\smallskip

Since $\nu_k=y_k+1/y_k$, potential singularities of $T_{1,1}$ may occur either at $v=\nu_k\pm{i\ov g}$  or at $v=w_k$. 
One can readily show that $T_{1,1}$ is regular at $v=w_k$  due to the auxiliary equations for the roots $w_k$. 
As to $v=\nu_k\pm{i\ov g}$, 
there are  two cases $y_k=x(\nu_k)$ and $y_k=1/x(\nu_k)$ to be discussed. In the first case one can easily see that a potential pole at $v=\nu_k+{i\ov g}$ cancels out due to the normalization factor  ${\mathscr N}_1$, and $T_{1,1}(v)$ is regular at $v=\nu_k+{i\ov g}$. Then, one can check that the potential pole in the normalization factor at $v=\nu_k-{i\ov g}$ cancels out due to the auxiliary Bethe equations for $\nu_k$. In the second case the normalization factor is regular for any $v$ but there still exists a potential pole at $v=\nu_k+{i\ov g}$. A careful analysis of the residue of $T_{1,1}$ at $v=\nu_k+{i\ov g}$ shows that it vanishes due to the auxiliary Bethe equations for $\nu_k$, and, therefore $T_{1,1}$ is regular at $v=\nu_k+{i\ov g}$. One can also show in a similar way that $1/T_{1,1}$ is regular at all these points.
Thus,  both $T_{1,1}$ and $1/T_{1,1}$ are regular for any value of $v$ related to the auxiliary roots $y_k\,, w_k$.  The same consideration shows that any asymptotic transfer matrix  $T_{a,1}$ and its inverse are also regular. Since $T_{a,s}$ are expressed via  $T_{a,1}$ by means of the BR formula, we conclude that any $T_{a,s}$ is regular for any value of $v$ related to the auxiliary roots $y_k\,, w_k$. One can also check that $1/T_{a,s}$  is regular for these values of $v$.

\smallskip

The regularity of $T_{a,s}$ and $1/T_{a,s}$ immediately implies that no Y-function can have either zero or pole or be equal to $-1$ at these $y_k\,, w_k$ related locations. Hence, we are to conclude that there are no exact Bethe equations for auxiliary roots, and they are just invisible in the TBA equations. Another evidence why this should be the case is that the number of auxiliary $y$-roots depends on the grading used, and, {\it e.g.}, an $N$-particle state from the $\su(2)$ sector has no auxiliary roots if one uses the $\su(2)$ grading and $2N$ $y$-roots if one uses the $\sl(2)$ grading. The asymptotic Y-functions are independent of the grading because as we show in appendix \ref{appT} the transfer matrices in the $\sl(2)$ and $\su(2)$ grading can be obtained one from another by a duality transformation, and therefore, they are just equal on the solutions of the corresponding auxiliary Bethe equations. Thus, it would be unclear why one should have some extra equations for the auxiliary roots if one uses the $\su(2)$ grading.

\smallskip

To summarize, a physical state is completely characterized by a set of momenta of particles it is made of,  and a set of auxiliary roots numbers $\KII_\a$ and $\KIII_\a$, while the auxiliary roots are definite functions of the momenta. As a result, the eigenvalues of transfer matrices and Y-functions are also determined by  the values of the momenta, and the location of the auxiliary roots is not reflected in their analytic properties.



\section{Implementation of the ${\rm PSU}(2,2|4)$ symmetry}\la{sec:psu}
Here we discuss the issue of the ${\rm PSU}(2,2|4)$ symmetry
in the mirror TBA approach. We start with the asymptotic Bethe Ansatz equations in the $\sl(2)$ grading \cite{BS}.
The main Bethe equations have the form
 \bea \label{BAmain} 
 &&1=e^{iJp_k}\prod_{l\neq
k}^{\KI}S_{\sl(2)}(u_k,u_l)\prod_{l=1}^{\KII_{-}}\frac{x_k^- -
y_l^{(-)} }{x_k^+-y_l^{(-)}}\sqrt{\frac{x_k^+}{x_k^-}}\, 
\prod_{l=1}^{\KII_{+}}\frac{x_k^- - y_l^{(+)}
}{x_k^+-y_l^{(+)}}\sqrt{\frac{x_k^+}{x_k^-}}\, .
\eea
These equations are supplied with auxiliary Bethe equations for the roots
$y^{(\alpha)}$ and $w^{(\alpha)}$, $\alpha=\pm $,
\bea
\label{BA1}
\prod_{i=1}^
{K^{\rm{I}}}\frac{y_k^{(\alpha)}-x^-_i}{y_k^{(\alpha)}-x^+_i}\sqrt{\frac{x^+_i}{x^
-_i}} &=&\prod_{i=1}^{K^{\rm{III}}_{\alpha}}
\frac{w_i^{(\alpha)}-\nu_k^{(\alpha)}-\frac{i}{g}}{w_i^{(\alpha)}-\nu_k^{(\alpha)}+\frac{i}{g}}\,  ,
\\ \label{BA2}
\prod_{i=1}^{\KII_{\alpha}}\frac{w_k^{(\alpha)}-\nu_i^{(\alpha)}+\frac{i}{g}}
{w_k^{(\alpha)}-\nu_i^{(\alpha)}-\frac{i}{g}}
&=&-\prod_{i=1}^{\KIII_{\alpha}}\frac{w_k^{(\a)}-w_i^{(\a)}+\frac{2i}{g}}{w_k^{(\a)}-w_i^{(\a)}-\frac{2i}{g}}\, .
\eea
Here we introduced a concise notation $\nu_k^{(\alpha)}=y_k^{(\alpha)}+\frac{1}{y_k^{(\alpha)}}$.
Solutions are therefore characterized by the
following five excitation numbers
$$
(\KIII_-,\KII_-,\KI,\KII_+,\KIII_+)\, .
$$
The number $\KI$ is a number of momentum-carrying particles, while $\KII_{\alpha}$ and 
$\KIII_{\alpha}$ give the weights of four ${\rm SU}(2)$ subgroups which represent a manifest symmetry of the string sigma model in the light-cone gauge.
The ${\rm SU}(4)$ weights $[q_1,p,q_2]$ and the spins $[s_1,s_2]$
of the corresponding excited state are \bea
\begin{array}{lll}
q_1=\KII_--2\KIII_- & ~~~~~~~~~~ & s_1=\KI-\KII_- \\
 p=J-\frac{1}{2}(\KII_-+\KII_+)+\KIII_-+\KIII_+ & & s_2=\KI-\KII_+
 \\
q_2=\KII_+-2\KIII_+ & &
\end{array}
\eea
Instead of weights of $\su(4)$
 one can use the weights $(J_1,J_2,J_3)$ of ${\rm SO}(6)$ and the relation between the two is 
 \bea J\equiv J_1=\frac{1}{2}(q_1+2p+q_2) \, , ~~~~~~~
J_2=\frac{1}{2}(q_1+q_2)\, , ~~~~~~
J_3=\frac{1}{2}(q_2-q_1) \, . \eea

Now we are ready to discuss the realization of  the   ${\rm PSU}(2,2|4)$
symmetry on asymptotic solutions. First, we recall that this symmetry can be consistently realized only on physical states \cite{BS}, {\it i.e.}  those which satisfy the level-matching condition. As can be seen from eqs.(\ref{BA1})
and ({\ref{BA2}}), such states might have a certain number of roots $\nu^{(\alpha)}$
and $w^{(\alpha)}$ located at  infinity\footnote{Obviously, the corresponding roots $y^{(\alpha)}$ can be located at either zero or infinity.}.
Second,  the states belonging to the same multiplet must have the one and the same anomalous dimension and canonical dimensions which might differ from each other by a half-integer only.
For the light-cone string sigma model the dispersion relation is 
\bea
E=J+\sum_{k=1}^{\KI}\sqrt{1+4g^2\sin^2\frac{p_k}{2}}\, .
\eea
For $g=0$ the last formula gives the canonical dimension of the corresponding gauge theory operator, while the difference $E-J-\KI$ corresponds to the anomalous dimension. Thus, adding 
particles with zero momentum changes the canonical dimension, but does not influence the
anomalous one and therefore should correspond to passing to a different member of a supersymmetry multiplet. Note that $p=0$ corresponds to $u=\infty$ in the standard 
$u$-plane parametrization of the momentum.  

\smallskip

What was said above  motivates the following treatment of the ${\rm PSU}(2,2|4)$
symmetry on the asymptotic solutions. We assume that every multiplet of ${\rm PSU}(2,2|4)$ has a {\it unique
regular} representative among the solutions of the Bethe ansatz
equations. Regularity means that all Bethe roots  
$(\vec{u},\vec{\nu}^{(\alpha)},\vec{w}^{(\alpha)})$ of the
corresponding solution  are finite. The regular representative\footnote{Which primary state is
regular depends on the grading chosen.} is a primary state of
the bosonic subgroup ${\rm SU}(2)\times {\rm SU}(2)\times {\rm
SU}(2)\times {\rm SU}(2)$ and it carries excitation numbers $\KI$,
$\KII_{\a}$ and $\KIII_{\a}$.  All the other states in the multiplet are obtained by adding roots at infinity. Obviously, the regular representative has a minimal number of Bethe roots.


\smallskip 

In spite of the fact that irregular states in a supermultiplet have excitation numbers different from the regular ones,  they must be nevertheless described by the same Bethe equations as for the regular state. From the main equation  (\ref{BAmain}),
one sees that adding a particle with $p=0$ does not influence this equation at all, since for this value of momentum $x^+/x^-=1$. The situation is different for $y$-roots: Adding a single $y=\infty$ root will leave eq.(\ref{BAmain}) invariant only 
if the original charge $J$ will be replaced as $J\to J-\frac{1}{2}$.  Oppositely, adding a single $y=0$ root requires a shift $J\to J+\frac{1}{2}$. Hence, there is a correlation between the number of irregular $y$-roots and the charge $J$ of the corresponding state. 
Finally, we note that adding any number of irregular $y$- or $w$-roots does not influence the auxiliary Bethe equations for a regular physical state.

\smallskip

It is of interest to determine the Bethe root content and weights (Dynkin labels) of the superconformal primary state corresponding to a given regular state. The structure of a generic superconformal multiplet is such that primary states with respect to the conformal group are in correspondence with states which are obtained by acting on the superconformal primary with 
all possible combinations of 16 Poincar\'e supercharges\footnote{Conformal supercharges annihilate a superconformal primary state.}. A superconformal primary state has obviously the lowest canonical dimension in the multiplet.  
To understand the action of the Poincar\'e supercharges, it is convenient to split them into two groups. We recall \cite{AFPZ} that in the uniform light-cone gauge $x_+=\tau$, where $\tau$ is the world-sheet time, the supercharges are naturally divided with respect to their dependence on 
the unphysical field $x_-$: they are either {\it kinematical} (independent of $x_-$) or {\it dynamical} (dependent on $x_-$). Kinematical supercharges do depend on $x_+$ and, for this reason, they do not commute with the world-sheet Hamiltonian 
${\rm H}=E-J$, while dynamical supercharges are independent of $x_+$ and commute with ${\rm H}$. In the Tables 1 and 2 we presented the weights of the kinematical and dynamical supercharges, respectively, as well as their action on the excitation numbers.  

\vskip 0.5cm

\begin{center}
{\small
 {
 \renewcommand{\arraystretch}{1.5}
\renewcommand{\tabcolsep}{0.2cm}
\begin{tabular}{||c|l|c|c|c|c|}
\hline Charge               & Weights                                                                   & $\Delta \KII_-$ & $\Delta \KII_+$ & $\Delta\KIII_-$ & $\Delta\KIII_+$\\
\hline $Q^3_{\a} $                 & $[0,-1,1]_{(\pm \frac{1}{2},0)}$                   &      $0_{+\frac{1}{2}}$, $2_{-\frac{1}{2}}$          &      $1_{+\frac{1}{2}} $, $1_{-\frac{1}{2}}$       &      $0_{+\frac{1}{2}}$, $1_{-\frac{1}{2}}$            &      $ 0_{\pm \frac{1}{2}}  $    \\
\hline $Q^4_{\a} $                 & $ [0,0,-1]_{(\pm\frac{1}{2},0)}$                  &      $0_{+\frac{1}{2}}$, $2_{-\frac{1}{2}}$            &     $1_{\pm \frac{1}{2}} $         &    $0_{+\frac{1}{2}}$,   $1_{-\frac{1}{2}}$           &     $ 1_{\pm \frac{1}{2}}  $        \\
\hline $\bar{Q}_{1\dot{a}}$  & $[-1,0,0]_{(0,\pm \frac{1}{2})}$                    &        $1_{\pm \frac{1}{2}} $                                  &       $0_{+\frac{1}{2}}$, $2_{-\frac{1}{2}}$        &     $ 1_{\pm \frac{1}{2}}  $          &       $0_{+\frac{1}{2}}$,   $1_{-\frac{1}{2}}$     \\
\hline $\bar{Q}_{2\dot{a}}$  & $ [1,-1,0]_{(0,\pm \frac{1}{2})}$                    &       $1_{+\frac{1}{2}} $, $1_{-\frac{1}{2}}$           &       $0_{+\frac{1}{2}}$, $2_{-\frac{1}{2}}$            &$ 0_{\pm \frac{1}{2}}  $ &     $0_{+\frac{1}{2}}$, $1_{-\frac{1}{2}}$        \\
 \hline
\end{tabular}
}
}

\vspace{0.5cm}
\parbox{13cm}
{\small Table 1. Kinematical Poincar\'e supercharges. These supercharges decrease $J$ by $-1/2$ and increase  $\KI$ by 1, they never decrease $\KII_{\a}$ and $\KIII_{\a}$. Here $\Delta \KII_{\a}$ and $\Delta \KIII_{\a}$ denote the change of  the corresponding excitation numbers under the action of a supercharge.}
\end{center}
\vspace{0.3cm}

Concerning kinematical supercharges, one can see that any such supersymmetry generator raises the number of zero momentum particles by one and, on the other hand, never lowers  all the other excitation numbers. 
More specifically, acting with one of these supercharges adds in total either one or three irregular $y$-roots, depending on a supercharge under consideration.
Thus, when applied to a superconformal primary state
these supercharges always add further irregular roots and, therefore, can never generate the corresponding regular state.  

\smallskip 

Note that we can easily determine the location of irregular $y$-roots created by the action of a kinematical supercharge. 
Since acting with such a charge leads to the shift $J\to J-\frac{1}{2}$, we conclude from our consideration of the main Bethe equation 
that for a charge generating a single $y$-root 
the latter must be at infinity, while for one generating three $y$-roots, two of them must be at infinity and one at zero. 

\vskip 0.5cm

\begin{center}
{\small
 {
 \renewcommand{\arraystretch}{1.5}
\renewcommand{\tabcolsep}{0.2cm}
\begin{tabular}{||c|l|c|c|c|c|}
\hline Charge               & Weights                                                                   & $\Delta\KII_-$ & $\Delta\KII_+$ & $\Delta\KIII_-$ & $\Delta\KIII_+$\\
\hline $Q^1_{\a} $                 & $[1,0,0]_{(\pm \frac{1}{2},0)}$                    &      $-1_{+\frac{1}{2}}$, $1_{-\frac{1}{2}}$          &      $0_{\pm \frac{1}{2}} $       &      $-1_{+\frac{1}{2}}$, $0_{-\frac{1}{2}}$            &      $ 0_{\pm \frac{1}{2}}  $    \\
\hline $Q^2_{\a} $                 & $ [-1,1,0]_{(\pm \frac{1}{2},0)}$                  &      $-1_{+\frac{1}{2}}$, $1_{-\frac{1}{2}}$            &     $0_{\pm \frac{1}{2}} $         &    $0_{+\frac{1}{2}}$,   $1_{-\frac{1}{2}}$           &     $ 0_{\pm \frac{1}{2}}  $        \\
\hline $\bar{Q}_{3\dot{a}}$  & $[0,1,-1]_{(0,\pm \frac{1}{2})}$                    &       $0_{\pm \frac{1}{2}} $           &       $-1_{+\frac{1}{2}}$, $1_{-\frac{1}{2}}$        &     $ 0_{\pm \frac{1}{2}}  $             &      $0_{+\frac{1}{2}}$,   $1_{-\frac{1}{2}}$       \\
\hline $\bar{Q}_{4\dot{a}}$  & $ [0,0,1]_{(0,\pm \frac{1}{2})}$                    &        $0_{\pm \frac{1}{2}} $          &       $-1_{+\frac{1}{2}}$, $1_{-\frac{1}{2}}$           & $ 0_{\pm \frac{1}{2}}  $ &      $-1_{+\frac{1}{2}}$, $0_{-\frac{1}{2}}$         \\
 \hline
\end{tabular}
}
}

\vspace{0.5cm}
\parbox{13cm}
{\small Table 2. Dynamical Poincar\'e supercharges. These supercharges do not change the value of $\KI$, but they raise both  $J$ and the canonical dimension by $1/2$. There exists four supercharges which lower
$\KII$ by 1.}
\end{center}
\vspace{0.3cm}

The situation with dynamical generators is a bit different. They do not change the value of $\KI$, as they simultaneously raise $E$ and $J$ by $1/2$. 
As one can see from Table 2, there are four supercharges which have non-negative excitation numbers, and, for the same reason as for kinematical supercharges, 
they cannot generate a regular state from its superconformal primary.  Any of the remaining four supercharges lowers $\KII$ by one and leaves intact or lowers 
$\KIII$ by one. 
It is these generators we are most interested in, because their application to a superconformal primary state decreases the number of irregular roots and, therefore, can produce a regular state.  
It is also clear that we should apply all these four generators to a superconformal primary to get the regular one, as in the opposite case, there remains a possibility to further lower the number of irregular roots. 
Denote by $E_{hws}$  and $J_{hws}$ the charges of a superconformal primary, which is the highest weight state of a supersymmetry multiplet, and by  $E_{reg}$  and $J_{reg}$ the charges of the corresponding regular state.
Hence, we conclude that\footnote{The reader can verify this formula for the case of the Konishi multiplet. We recall that the Konishi superconformal primary operator has the canonical dimension $\Delta=2$ and $J=0$. It has a regular descendent in the $\sl(2)$ sector
which has $\Delta=4$ and $J=2$.}
\bea
\label{EJhws}
E_{hws}=E_{reg}-2\, ,~~~~~~ J_{hws}=J_{reg}-2\, .
\eea
Also, the relationship between the corresponding excitation numbers is 
\bea
\KI_{reg}=\KI_{hws}\, ,~~~~~
\KII_{\a,reg}= \KII_{\a, hws}-2 \, , ~~~~~ \KIII_{\a,reg}= \KIII_{\a, hws}-1\eea
for both $\a$.  Our discussion of the main Bethe equation together with the relation $J_{hws}=J_{reg}-2$ implies that four irregular $y$-roots which distinguish a superconformal primary from its regular state must all be located at infinity.

\smallskip

Similar to what has been done for kinematical generators, we can now establish locations of irregular $y$-roots created by the action of dynamical generators. Since application of dynamical generators shifts $J\to J+\frac{1}{2}$,
four of them generate  four roots at $y=0$, while the other four remove four roots at $y=\infty$. This leads to the following description of an arbitrary state generated by a superconformal primary $|{\rm hws}\rangle$
\bea
\prod (Q_{-\infty}^{d})^{n_{\infty}^d}
(Q_{+0}^{d})^{n_{0}^d}(Q_{+\infty}^{k})^{n_{\infty}^k}(Q_{+2\infty,+0}^{k})^{n_{\infty,0}^k}|{\rm
hws}\rangle\, ,\eea
where the integers $n_{\infty}^d,\dots, n_{\infty,0}^k$ can take any value
from zero to four. Here the upper subscript $``d"$ or $``k"$ in the definition of a supercharge $Q$ specifies if it is dynamical or kinematical, respectively.
The subscript of $Q$ points the location and the number of $y$-roots with $+$ and $-$ sign signifying if the corresponding root is added or removed, respectively.
Taking into account how each supercharge
increases or decreases the value of $J$, one finds that such a state
has \bea
J=J_{hws}+\frac{1}{2}(n_{\infty}^d+n_{0}^d-n_{\infty}^k-n_{\infty,0}^k)\,
 \eea
and its energy is 
\bea E=E_{hws}
+\frac{1}{2}(n_{\infty}^d+n_{0}^d+n_{\infty}^k+n_{\infty,0}^k)\, .
\eea
Also, this state has the following number of main particles
\bea
 \KI =\KI_{reg}+n_{\infty}^k+n_{\infty,0}^k \, .\eea
As a check, we get
 \bea
 E-J=E_{hws}&-&J_{hws}+n_{\infty}^k+n_{\infty,0}^k=\\ \nonumber
 &=&E_{reg}-J_{reg}+ n_{\infty}^k+n_{\infty,0}^k=K^I_{reg}+n_{\infty}^k+n_{\infty,0}^k=\KI\, ,\eea
{\it i.e.} as expected for the free dispersion relation for any member of the multiplet.
Further, we see that for the state under consideration a number of {\it irregular} roots, denoted by ${\cal K}^{\rm II}$, is   
 \bea
 {\cal K}^{\rm II}_{\infty}&=&4-n_{\infty}^d+n_{\infty}^k+2n_{\infty,0}^k \, ,\\
{\cal K}^{\rm II}_{0}&=&n_{0}^d+n_{\infty,0}^k\, \eea 
so that the total number of $y$-roots is $\KII=\KII_{reg}+{\cal K}^{\rm II}_{0}+ {\cal K}^{\rm II}_{\infty}$.
This allows  one to express $J$ in terms of $J_{reg}$ and 
the number of irregular $y$-roots
 \bea
 \label{JvsJhws}
J=J_{hws}+2+\sfrac{1}{2}({\cal K}^{\rm II}_{0}-{\cal K}^{\rm II}_{\infty})=J_{reg}+\sfrac{1}{2}({\cal K}^{\rm II}_{0}-{\cal K}^{\rm II}_{\infty})\, .\eea
Obviously, this formula is in complete agreement with the fact that the state we consider must have the same Bethe equations as the corresponding regular state.

\smallskip

Now we turn our attention to the large $J$ asymptotic solution of the mirror TBA equations.  
The asymptotic $Y_Q$-functions for a state of charge $J$ and excitation numbers $\KI$, $\KII$ and $\KIII$ are given by eqs.(\ref{YQasympt}). 
It is important to realize that this expression for $Y_Q$ is valid not only for a regular state, but for any state 
in the corresponding superconformal multiplet. As we now show, all states in a multiplet have in fact the one and same asymptotic $Y_Q$-functions. 
First, from the explicit expression (\ref{eqn;FullEignvalue1}) for transfer matrices $T_{Q,\pm 1}$, one can see that they remain unchanged if one adds any number of 
infinite roots $u$ or $w^{(\a)}$. Second, if a state has  a number   of $y$-roots ${\cal K}^{\rm II}_{0}$   at zero and a number of $y$-roots ${\cal K}^{\rm II}_{\infty}$ 
at infinity, then the expression for transfer matrices shows that they  are related to the transfer matrices of the corresponding regular state 
in a very simple fashion
\bea
T_{Q,+1}T_{Q,-1}=\Big(\frac{x^+}{x^-}\Big)^{\sfrac{1}{2}({\cal K}^{\rm II}_{\infty}-{\cal K}^{\rm II}_{0}) }T_{Q,+1}^{reg}T_{Q,-1}^{reg}\, .
\eea
Accordingly, the $Y_Q$-functions of this state can be written as 
\bea
Y_Q^o=\Big(\frac{x^+}{x^-}\Big)^{J-J_{reg}}\Big(\frac{x^+}{x^-}\Big)^{\sfrac{1}{2}({\cal K}^{\rm II}_{\infty}-{\cal K}^{\rm II}_{0}) }Y_Q^{o,reg}=Y_Q^{o,reg}\, ,\eea
where in the last formula we used eq.(\ref{JvsJhws}). Thus, for all states in a multiplet the $Y_Q^o$-functions are the same as for the regular state. 
Since the knowledge of $Y_Q^o$ allows one to find all the other asymptotic Y-functions, we conclude that
all states in a superconformal multiplet must share the one and the same asymptotic solution. Moreover, since the excited states TBA equations are engineered
by using the analytic properties of the asymptotic Y-functions, we conclude that these equations are constructed not for a particular state but rather for a whole supersymmetry multiplet. This implies that the ${\rm PSU}(2,2|4)$ symmetry is in some sense built in into the mirror TBA approach, in a similar way as it is for 
the asymptotic Bethe ansatz.

\smallskip

Now we point out an interesting interpretation of the TBA length parameter $L_{TBA}$. Applying all kinematical generators to a highest weight state of charge $J_{hws}$
, in a generic situation $J>3$, we obtain a state of the lowest $J$-charge, $J=J_{hws}-4$.  Oppositely, acting on $|{\rm hws}\rangle$
with all dynamical generators produces a state of the highest $J$-charge, $J=J_{hws}+4$. Thus, the $J$-charge of any state in a generic multiplet obeys inequalities 
\bea J_{hws}-4\leq J\leq
J_{hws}+4=J_{reg}+2\, . \eea
As was found in  \cite{AFS09}, the length parameter $L_{TBA}$ must be related to the $J$-charge  of a regular state as 
$$L_{TBA}=J_{reg}+2\,.$$  
Thus, $L_{TBA}$ simply coincides with the maximal $J$-charge in a supersymmetry multiplet. 

\smallskip

One remark is in oder. A generic supersymmetry multiplet is obtained by free action of 16 Poincar\'e supercharges on a highest weight state (a state of the lowest dimension in the multiplet) and arising representations of the maximal bosonic subgroup have Dynkin labels which are obtained by adding the weights of the corresponding supercharges to 
Dynkin labels of the highest weight state.  In the case where the resulting labels turn out to be  negative (non-generic multiplets), special rules must be applied to find the corresponding Dynkin content\footnote{For ${\rm PSU}(2,2|4)$ these rules can be found, for instance, in \cite{Dolan:2002zh}. }.   The Konishi multiplet, which has $J_{hws}=0$, provides an example of a non-generic multiplet; it has $2^{16}$ states which are organized in 532 representations of the maximal 
bosonic subgroup of the conformal group.  Nevertheless, certain formulas we discussed in this section remain valid for the Konishi multiplet, for instance,
the relations  (\ref{EJhws}), and also $L_{TBA}=4$.

\section*{Acknowledgements}
We are grateful to Niklas Beisert for discussions concerning superconformal symmetry.
G.A. acknowledges support by the Netherlands Organization for Scientific Research (NWO) under the VICI grant 680-47-602.
The work of S.F. was supported in part by the Science Foundation Ireland under Grant 09/RFP/PHY2142. 

\section{Appendix}
\subsection{ Simplified TBA equations and Y-system}\la{appYsys}

The simplified TBA equations for the ground state derived in \cite{AF09b,AF09d} can be written in the following form\footnote{We set the regularization parameters $h_\a$ to 0, where the parameter $\a$ takes the values $\pm$. In our previous papers $\a=1,2$.} 

\noindent
$\bullet$\  $Q=1$-particle
\bea
\nonumber
\log Y_{1}(v)&=&
\log\left(1-{1\ov Y_{-}^{(-)}} \right)\left(1-{1\ov Y_{-}^{(+)}} \right)Y_2\, \hat{\star}\, s- \log(1 + Y_{2})\star s \\\nonumber
&-& \log\left(1-{1 \ov Y_{-}^{(-)}} \right)\left(1-{1 \ov Y_{+}^{(-)}}\right)\left(1-{1 \ov Y_{-}^{(+)}} \right)\left(1-{1 \ov Y_{+}^{(+)}}\right)
Y_2^2\, \hat{\star}\, \check{K}\,\check{\star}\,  s 
\\\la{YforQ1}
&-& \log\left(1+Y_{Q} \right)\star \big( 2\check{K}_Q^\Sigma +
\check{K}_Q  +\check{K}_{Q-2}\big)\,\check{\star}\,  s +\log{Y_1}\star \check{K}_1\,\check{\star}\,  s-L\, \check{\cal E}\,\check{\star}\, s\,.~~~~~~~
\eea
Here and in what follows we use the definitions and conventions  from
\cite{AFS09}. 

It has been found \cite{AFS09} that for all the excited states analyzed in the TBA approach  the length parameter $L$ is related to the charge $J$ carried by a string state as $L=J+2\,.$
We argue in section \ref{sec:psu} that the relation between length and charge is universal and holds for any excited state with regular Bethe roots.

Assuming that   $|v|\le2$ and acting on \eqref{YforQ1} by the operator $s^{-1}$, one derives the first Y-system equation
\bea\la{YQ1}
{Y_{1}^{+}\, Y_{1}^{-}\ov Y_{2}} &=& {\left(1 -  {1\ov Y_{-}^{(-)}} \right)
\left(1 -  {1\ov Y_{-}^{(+)}} \right)\ov  1+Y_2 }\, ,\eea
where we use the notation $f^\pm=f(v\pm{i\ov g}\mp i0)$ and take into account that for $|v|\le2$ only the first line in  \eqref{YforQ1} contributes. 

\bigskip
 \noindent
$\bullet$\  $Q$-particles 
\bea
\log Y_{Q+1}&=&\log{\left(1 +  {1\ov Y_{Q|vw}^{(-)}} \right)\left(1 +  {1\ov Y_{Q|vw}^{(+)}} \right)\ov (1 +  {1\ov Y_{Q} })(1 +  {1\ov Y_{Q+2} }) }\star s\la{YforQ}
\,,\quad Q\ge 1\,.~~~~~~~
\eea
This equations lead to the following Y-system equations valid for any $v$
\bea\la{YQQ}
{Y_{Q+1}^{+}\, Y_{Q+1}^{-}\ov Y_{Q}Y_{Q+2}} &=& {\left(1 +  {1\ov Y_{Q|vw}^{(-)}} \right)
\left(1 +  {1\ov Y_{Q|vw}^{(+)}} \right)\ov \left( 1+Y_{Q} \right)\left( 1+Y_{Q+2} \right) } \,.
\eea
 
\bigskip
 \noindent
$\bullet$\   $y$-particles, $\a=\pm$  
\bea
\la{Yfory1}
\log {Y_+^{(\a)} \ov Y_-^{(\a)} }(v)&=&\log(1 +  Y_{Q})\star K_{Qy}\,,~~~~~~~ \\
\la{Yfory2}
\log {Y_+^{(\a)}  Y_-^{(\a)} }(v) &=&2\log{1 +  Y_{1|vw}^{(\a)} \ov 1 +  Y_{1|w}^{(\a)} }\star s
\\\nonumber
&-&
\log\left(1+Y_Q \right)\star K_Q+ 2 \log(1 +  Y_{Q})\star K_{xv}^{Q1} \star s\,.~~~~
\eea
The AdS/CFT Y-system is incomplete and contains equations only for $Y_-^{(\a)} $-functions.
Their derivation is not straightforward  and can be found in \cite{AF09b}. 
Assuming that $Y_\pm^{(\a)} $-functions are analytic in the vicinity of 
the interval $|v|\le 2$, one gets
\bea\la{Ym}
Y_{-}^{(\a)+}\,Y_{-}^{(\a)-}  &=& {1+ Y_{1|vw}^{(\a)} \ov1+ Y_{1|w}^{(\a)}} \, {1\ov 1+Y_1 }
\,.
\eea
This assumption is compatible with  the kernel $K_{Qy}$ where the cut is chosen to be for $|v|\ge2$. One can, however, choose the cut to be for  $|v|\le2$, which is in fact more natural if one thinks about Y-functions as being defined on a $z$-torus. To discuss the asymptotic solution it is easier to use the cut $|v|\ge2$.

\bigskip
 \noindent
$\bullet$\ $M|vw$-strings: $\ M\ge 1\ $, $Y_{0|vw}=0$
\bea\la{Yforvw}
\hspace{-0.3cm}\log Y_{M|vw}^{(\a)} (v)&=&- \log(1 +  Y_{M+1})\star s~~~~~\\\nonumber
&+& \log(1 +  Y_{M-1|vw}^{(\a)}  )(1 +  Y_{M+1|vw}^{(\a)} )\star s+\delta_{M1}  \log{1-Y_-^{(\a)} \ov 1-Y_+^{(\a)} }\hstar s\,,~~~~~
\eea
and the Y-system equations
\bea
\la{Yvw1}
Y_{1|vw}^{(\a)+}\, Y_{1|vw}^{(\a)-} &=& { 1+Y_{2|vw}^{(\a)} \ov 1+Y_{2}} \, {1-Y_-^{(\a)}\ov 1-Y_+^{(\a)}} \\
\la{YvwM}
Y_{M|vw}^{(\a)+}\, Y_{M|vw}^{(\a)-} &=& \left( 1+Y_{M-1|vw}^{(\a)} \right)\left( 1+Y_{M+1|vw}^{(\a)} \right)\, {1\ov 1+Y_{M+1}}\,,  \quad M\ge 2\,,~~~~~~~
\eea

\bigskip
 \noindent
$\bullet$\ $M|w$-strings: $\ M\ge 1\ $, $Y_{0|w}=0$
\bea\la{Yforw}
\log Y_{M|w}^{(\a)} =  \log(1 +  Y_{M-1|w}^{(\a)} )(1 +  Y_{M+1|w}^{(\a)} )\star s
+\delta_{M1}\, \log{1-{1\ov Y_-^{(\a)} }\ov 1-{1\ov Y_+^{(\a)} } }\hstar s\,.~~~~~
\eea
and the Y-system equations
\bea
\la{Yw1}
Y_{1|w}^{(\a)+}\,Y_{1|w}^{(\a)-} &=& \left( 1+Y_{2|w}^{(\a)} \right){1-{1\ov Y_-^{(\a)}}\ov 1-{1\ov Y_+^{(\a)}} }\\
\la{YwM}
Y_{M|w}^{(\a)+}\,Y_{M|w}^{(\a)-} &=& \left( 1+Y_{M-1|w}^{(\a)} \right)\left( 1+Y_{M+1|w}^{(\a)} \right)\,,  \quad  M\ge 2\,, 
\eea
Let us stress again that the equations above are valid only for $|v|\le2$. For other values of $v$ one should use an analytic continuation, and the resulting equations depend on it.


\subsection{ Duality transformation and transfer matrices }\la{appT}
The duality transformation of the $y$-roots can be implemented on the level of transfer matrices and used to obtain explicit expressions for 
$T_{a,1}$ and $T_{1,s}$ in the $\sl(2)$ and $\su(2)$ gradings. This consideration seems to give support to the statement that auxiliary roots do not play any role for T-functions because their number and their values change under a duality transformation. 

\smallskip

The eigenvalues of the transfer matrix $T_{a,1}$ in the $\sl(2)$-grading conjectured in \cite{B06} were obtained in \cite{ALST} by using previously found 
scattering matrices for string bound states. The transfer matrix $T_{a,1}$ depends on the rapidities $u_1,\ldots, u_N$ of $N\equiv \KI$ physical particles and on the rapidity 
$v$ of an auxiliary particle that transforms in the bound state representation $(a,1)$ of $\psu(2|2)$.   Eigenvalues of $T_{a,1}$  
can be parametrized by a set $(y^{(+)},w^{(+)})$
of auxiliary roots which satisfy the set (\ref{BA1}), (\ref{BA2}) of auxiliary Bethe equations. The Y-functions also involve the transfer matrices $T_{a,-1}$, which have the same structure as $T_{a,1}$ with the replacement  $(y^{(+)},w^{(+)})$ for  $(y^{(-)},w^{(-)})$. Thus, in what follows we consider  $T_{a,1}$ only
and, for the sake of simplicity\footnote{We will also not distinguished between a transfer matrix and its eigenvalues.}, denote the auxiliary roots as    $(y,w)$.
The transfer matrix $T_{a,1}$ in the $\sl(2)$ grading reads
\begin{eqnarray}\label{eqn;FullEignvalue1}
&&T_{a,1}^{\sl(2)}(v)=\prod_{i=1}^{K^{\rm{II}}}{\textstyle{\frac{y_i-x^-}
{y_i-x^+}\sqrt{\frac{x^+}{x^-}} \, }}\left[1+
\prod_{i=1}^{K^{\rm{II}}}{\textstyle{
\frac{v-\nu_i+\frac{i}{g}a}{v-\nu_i-\frac{i}{g}a}}}\prod_{i=1}^{K^{\rm{I}}}
{\textstyle{\left[\frac{(x^--x^-_i)(1-x^-
x^+_i)}{(x^+-x^-_i)(1-x^+
x^+_i)}\frac{x^+}{x^-}  \right]}}  \right.\\
&&{\textstyle{+}}
\sum_{k=1}^{a-1}\prod_{i=1}^{K^{\rm{II}}}{\textstyle{
\frac{v-\nu_i+\frac{i}{g}a}{v-\nu_i+\frac{i}{g}(a-2k)}}} \Big[
\prod_{i=1}^{K^{\rm{I}}}{\textstyle{\frac{x(v+(a-2k)\frac{i}{g})-x_i^-}{x(v+(a-2k)\frac{i}{g})-x_i^+}}}+
\prod_{i=1}^{K^{\rm{I}}}{\textstyle{\frac{1-x(v+(a-2k)\frac{i}{g})x_i^-}{1-x(v+(a-2k)\frac{i}{g})x_i^+}
}}\Big]\prod_{i=1}^{K^{\rm{I}}}{\textstyle{\frac{x^+-x_i^+}{x^+-x_i^-}\frac{v-v_i-(2k+1-a)\frac{i}{g}}{v-v_i+(a-1)\frac{i}{g}
}}}\nonumber\\
&& -\sum_{k=0}^{a-1}\prod_{i=1}^{K^{\rm{II}}} {\textstyle{
\frac{v-\nu_i+\frac{i}{g}a}{v-\nu_i+\frac{i}{g}(a-2k)}}}\prod_{i=1}^
{K^{\rm{I}}}{\textstyle{\frac{x^+-x^+_i}{x^+-x^-_i}\sqrt{\frac{x^-_i}{x^
+_i}} \frac{v-v_i-(2k+1-a)\frac{i}{g}}{v-v_i+(a-1)\frac{i}{g}
}}}\prod_{i=1}^{K^{\rm{III}}}{\textstyle{\frac{w_i-v
+\frac{i(2k-1-a)}{g}}{w_i-v+\frac{i(2k+1-a)}{g}} }} \nonumber\\
&&\left. -\sum_{k=0}^{a-1}\prod_{i=1}^{K^{\rm{II}}} {\textstyle{
\frac{v-\nu_i+\frac{i}{g}a}{v-\nu_i+\frac{i
}{g}(a-2k-2)}}}\prod_{i=1}^
{K^{\rm{I}}}{\textstyle{\frac{x^+-x^+_i}{x^+-x^-_i}\sqrt{\frac{x^-_i}{x^
+_i}} \frac{v-v_i-(2k+1-a)\frac{i}{g}}{v-v_i+(a-1)\frac{i}{g}
}}}\prod_{i=1}^{K^{\rm{III}}}{\textstyle{\frac{w_i-v+\frac{i}{g}(2k+3-a)}{
w_i-v+\frac{i}{g}(2k+1-a)}}}\right]. \nonumber
\end{eqnarray}
In this formula
$$
v=x^++\frac{1}{x^+}-\frac{i}{g}a=x^-+\frac{1}{x^-}+\frac{i}{g}a\, .
$$
The variable $v$ takes values in the mirror theory $v$-plane, so that $x^{\pm}=x(v\pm \frac{i}{g}a)$ with $x(v)$ being the mirror theory $x$-function. 
Similarly, $x^{\pm}_j=x_s(u_j\pm \frac{i}{g})$, where $x_s$ is the string theory $x$-function. The overall factor 
\bea
\mathscr{N}_a(v)=\prod_{i=1}^{K^{\rm{II}}}{\textstyle{\frac{y_i-x^-}
{y_i-x^+}\sqrt{\frac{x^+}{x^-}} \, }}\eea
satisfies $\mathscr{N}^+_a\mathscr{N}^-_a=\mathscr{N}_{a-1}\mathscr{N}_{a+1}$ and is, therefore, a gauge transformation which drops from the auxiliary Y-functions.

\smallskip

Concerning the structure of $T_{a,1}^{\sl(2)}$, 
we point out that the unity occurring in the first line of  (\ref{eqn;FullEignvalue1}) can be considered as
coming from the first product  (in the square brackets) in the second line with $k=0$, while
the second term in the first line can be considered as coming from
the first product in the second line with $k=a$.

\smallskip

To discuss the duality transformation, see e.g. \cite{Essler, BS,B06} and reference therein,\footnote{The derivation of the dual form of the Bethe equations basically repeats the one performed in section 3 of \cite{B06}, and is presented here just for completeness.} we introduce the following polynomial in the variable $y$ of degree\footnote{To simplify the derivation, we assume that the level-matching condition is not imposed. We impose the level-matching condition after the duality transformation is done.}
$\KI+2\KIII$ \bea P(y)&=&\pI
(y-x_i^-)\spm\pIII y\Big(w_i-y-\frac{1}{y}+\IG\Big)\nonumber\\
&-&\pI (y-x_i^+)\pIII y\Big(w_i-y-\frac{1}{y}-\IG\Big) \, .
\label{pol}
\eea 
This polynomial has $\KII$ roots $y_i$ that are
solutions of the Bethe equations (\ref{BA1}). Therefore, this polynomial can be
written in the form 
\bea P(y)=c\pII(y-y_i)\pIIt(y-\tilde{y}_i) \,.\eea 
Here $\widetilde{K}^{\rm II}=\KI-\KII+2\KIII$ is the number of
the {\it dual} roots $\tilde{y}_i$. Thus, the ratio \bea
R(y)=\frac{P(y)}{\pII(y-y_i)\pIIt(y-\tilde{y}_i)}=c \eea is a
constant independent of $y$. As a result, one has $R(a)=R(b)$ for any $a$ and $b$,
that is \bea\label{maindual} \pII
\frac{y_i-a}{y_i-b}&=&\frac{P(a)}{P(b)}\pIIt\frac{\tilde{y}_i-b}{\tilde{y}_i-a}\,
.\eea
 In particular, \bea R(x^+)=R(x^-)\,
,~~~~~~~~~~~R(1/x^+)=R(1/x^-)\, , \eea yielding  \bea \hspace{-0.3cm}\pII
\frac{y_i-x^-}{y_i-x^+}=\frac{P(x^-)}{P(x^+)}\pIIt\frac{\tilde{y}_i-x^+}{\tilde{y}_i-x^-}\,
,~~~~
\pII\frac{y_i-\frac{1}{x^-}}{y_i-\frac{1}{x^+}}=
\frac{P(\frac{1}{x^-})}{P(\frac{1}{x^+})}\pIIt\frac{\tilde{y}_i-\frac{1}{x^+}}{\tilde{y}_i-\frac{1}{x^-}}\,
.
 \eea

We start with showing how the auxiliary Bethe equations (\ref{BA1}) and (\ref{BA2}) are dualized. By construction, $\tilde{y}_k$ is a root of $P(y)$, {\it i.e. }
 $P(\tilde{y}_k)=0$, which is nothing else but the Bethe equations for $\tilde{y}_k$:
  \bea \prod_{i=1}^
{K^{\rm{I}}}\frac{\tilde{y}_k-x^-_i}{\tilde{y}_k-x^+_i}\sqrt{\frac{x^+_i}{x^
-_i}} =\prod_{i=1}^{K^{\rm{III}}}\frac{w_i-\tilde{\nu}_k
-\frac{i}{g}}{w_i-\tilde{\nu}_k+\frac{i}{g}}\, , \eea 
 where $k=1,\ldots, \widetilde{K}^{\rm II}$.
Next, introducing $x^{\pm}(w)$ 
\bea
w\pm \frac{i}{g}=x^{\pm}(w)+\frac{1}{x^{\pm}(w)}\, ,
\eea
we factorize 
\bea
w_k-y-\frac{1}{y}\pm\frac{i}{g}=(x^{\pm}(w_k)-y)\Big(1-\frac{1}{y x^{\pm}(w_k)}\Big)\, .
\eea
Thus, 
\bea
\prod_{i=1}^{\KII}\frac{w_k-\nu_i+\frac{i}{g}}{w_k-\nu_i-\frac{i}{g}}&=&\prod_{i=1}^{\KII}\frac{(y_i-x^+(w_k))\Big(y_i-\frac{1}{x^+(w_k)}\Big)}
{(y_i-x^-(w_k))\Big(y_i-\frac{1}{x^-(w_k)}\Big)}=\\
\nonumber
&&\hspace{2cm}=\frac{P(x^+(w_k))P\left(\frac{1}{x^+(w_k)}\right)}{P(x^-(w_k))P\left(\frac{1}{x^-(w_k)}\right)} 
\prod_{i=1}^{\tilde{K}^{\rm{II}} }\frac{w_k-\tilde{\nu}_i-\frac{i}{g}}{w_k-\tilde{\nu}_i+\frac{i}{g}}\, .
 \eea 
 Therefore, the Bethe equations (\ref{BA2}) acquire the form
 \bea
 \label{BA2inter}
 \prod_{i=1}^{\tilde{K}^{\rm{II}} }\frac{w_k-\tilde{\nu}_i+\frac{i}{g}}{w_k-\tilde{\nu}_i-\frac{i}{g}}=-
 \frac{P(x^+(w_k))P\left(\frac{1}{x^+(w_k)}\right)}{P(x^-(w_k))P\left(\frac{1}{x^-(w_k)}\right)} \prod_{i=1}^{\KIII}
 \frac{w_k-w_i-\frac{2i}{g}}{w_k-w_i+\frac{2i}{g}}\, .\eea
 By using eq.(\ref{pol}), it is easy to compute
\bea
\nonumber
 \frac{P(x^+(w_k))P\left(\frac{1}{x^+(w_k)}\right)}{P(x^-(w_k))P\left(\frac{1}{x^-(w_k)}\right)} &=&
 \prod_{i=1}^{\KI}\frac{(x^+(w)-x^+_i)\Big(1-\frac{1}{x^+_ix^+(w_k)}\Big)}{(x^-(w)-x^-_i)\Big(1-\frac{1}{x^-_ix^-(w_k)}\Big)}
\prod_{i=1}^{\KIII}\left(\frac{w_k-w_i+\frac{2i}{g}}{w_k-w_i-\frac{2i}{g}}\right)^2=\nonumber \\
&&\hspace{3cm}=\prod_{i=1}^{\KIII}\left(\frac{w_k-w_i+\frac{2i}{g}}{w_k-w_i-\frac{2i}{g}}\right)^2\, .\eea
 Substituting this formula into (\ref{BA2inter}), we obtain the dualized auxiliary Bethe equations
 \bea
 \prod_{i=1}^{\tilde{K}^{\rm{II}} }\frac{w_k-\tilde{\nu}_i+\frac{i}{g}}{w_k-\tilde{\nu}_i-\frac{i}{g}}= \prod_{i=1}^{\KIII}
 \frac{w_k-w_i+\frac{2i}{g}}{w_k-w_i-\frac{2i}{g}}\, . \eea
 
 Let us now discuss the duality transformation of $T_{a,1}^{\sl(2)}$. For the simplest case of $T_{1,1}$ the transformation was used in section 5 of \cite{B06}. 
To perform this transformation, we combine the term with
$k=0$ from the third line with the unit in the first line,
also, the term with $k=a-1$ from the forth line will be combined
with the second term in the first line; in the remaining sum in
the forth line we make the change of the summation index $k\to
k+1$. This rearrangement yields
\begin{eqnarray}\label{eqn;FullEignvalue3}
&&T_{a,1}^{\sl(2)}(v)=\prod_{i=1}^{K^{\rm{II}}}{\textstyle{\frac{y_i-x^-}
{y_i-x^+}\sqrt{\frac{x^+}{x^-}} \,
}}\left[1-\prod_{i=1}^{K^{\rm{I}}}{\textstyle{\frac{x^+-x^+_i}{x^+-x^-_i}\sqrt{\frac{x^-_i}{x^
+_i}} }}\pIII
{\textstyle{\frac{w_i-v-\frac{i}{g}(a+1)}{w_i-v-\frac{i}{g}(a-1)}}}
\right. \\
&& +\prod_{i=1}^{K^{\rm{II}}}{\textstyle{
\frac{v-\nu_i+\frac{i}{g}a}{v-\nu_i-\frac{i}{g}a}}}\Big[\prod_{i=1}^{K^{\rm{I}}}
{\textstyle{\frac{(x^--x^-_i)(1-x^- x^+_i)}{(x^+-x^-_i)(1-x^+
x^+_i)}\frac{x^+}{x^-}}} -\prod_{i=1}^
{K^{\rm{I}}}{\textstyle{\frac{x^+-x^+_i}{x^+-x^-_i}\sqrt{\frac{x^-_i}{x^
+_i}} \frac{v-v_i-(a-1)\frac{i}{g}}{v-v_i+(a-1)\frac{i}{g}
}}}\prod_{i=1}^{K^{\rm{III}}}{\textstyle{\frac{w_i-v
+\frac{i(a+1)}{g}}{w_i-v+\frac{i(a-1)}{g}} }} \Big]\nonumber\\
\nonumber &&+\sum_{k=1}^{a-1}\left[
\prod_{i=1}^{K^{\rm{II}}}{\textstyle{
\frac{v-\nu_i+\frac{i}{g}a}{v-\nu_i+\frac{i}{g}(a-2k)}}} \Big[
\prod_{i=1}^{K^{\rm{I}}}{\textstyle{\frac{x(v+(a-2k)\frac{i}{g})-x_i^-}{x(v+(a-2k)\frac{i}{g})-x_i^+}}}
{\textstyle{\frac{x^+-x_i^+}{x^+-x_i^-}\frac{v-v_i+(a-2k-1)\frac{i}{g}}{v-v_i+(a-1)\frac{i}{g}
}}}+ \right.\\
\nonumber &&~~~~~~+
\prod_{i=1}^{K^{\rm{I}}}{\textstyle{\frac{1-x(v+(a-2k)\frac{i}{g})x_i^-}{1-x(v+(a-2k)\frac{i}{g})x_i^+}
}}{\textstyle{\frac{x^+-x_i^+}{x^+-x_i^-}\frac{v-v_i+(a-2k-1)
\frac{i}{g}}{v-v_i+(a-1)\frac{i}{g} }}}\\
\nonumber &&~~~~~~-\prod_{i=1}^
{K^{\rm{I}}}{\textstyle{\frac{x^+-x^+_i}{x^+-x^-_i}\sqrt{\frac{x^-_i}{x^
+_i}} \frac{v-v_i+(a-2k-1)\frac{i}{g}}{v-v_i+(a-1)\frac{i}{g}
}}}\prod_{i=1}^{K^{\rm{III}}}{\textstyle{\frac{w_i-v
+\frac{i(2k-1-a)}{g}}{w_i-v+\frac{i(2k+1-a)}{g}} }}\\
\nonumber &&~~~~~~\left.-\prod_{i=1}^
{K^{\rm{I}}}{\textstyle{\frac{x^+-x^+_i}{x^+-x^-_i}\sqrt{\frac{x^-_i}{x^
+_i}} \frac{v-v_i+(a-2k+1)\frac{i}{g}}{v-v_i+(a-1)\frac{i}{g}
}}}\prod_{i=1}^{K^{\rm{III}}}{\textstyle{\frac{w_i-v
+\frac{i(2k+1-a)}{g}}{w_i-v+\frac{i(2k-1-a)}{g}} }} \Big]\right]\,
.
\end{eqnarray}
By using the polynomial $P$ defined in  (\ref{pol}) and performing tedious but straightforward computation, one can show that the last formula can be written in the following more compact form 

\bea
\label{Ta1sl2}
T_{a,1}^{\sl(2)}(v)&=&{\textstyle{P(x^+)}}\displaystyle{\prod_{i=1}^{\KII} \textstyle{\frac{y_i-x^-}{y_i-x^+}\sqrt{\frac{x^+}{x^-}}}}\,
\prod_{i=1}^{\KI}{\textstyle\frac{ 1}{x^+-x_i^-}\sqrt{\frac{x^-_i}{x^+_i}}}\,  \displaystyle{\prod_{i=1}^{\KIII}}{\textstyle\frac{1}{x^+(w_i-v-\frac{i}{g}(a-1))}} \\
\nonumber
&-&{\textstyle{P\left(\frac{1}{x^-}\right)}}\displaystyle{\prod_{i=1}^{\KII}{\textstyle\frac{y_i-\frac{1}{x^+}}{y_i-\frac{1}{x^-}}\sqrt{\frac{x^+}{x^-}}}
\prod_{i=1}^{\KI}{\textstyle \frac{x^--x_i^-}{(x^+-x_i^-)(\frac{1}{x^+}-x_i^+)}}
}\displaystyle{\prod_{i=1}^{\KIII}{\textstyle\frac{x^-}{w_i-v+\frac{i}{g}(a-1)}}}\\
\nonumber
&-&\sum_{k=1}^{a-1}{\textstyle{P\Big(\frac{1}{x(v+\frac{i}{g}(a-2k))}\Big)P\Big(x(v+\frac{i}{g}(a-2k))\Big)}}
\displaystyle{\prod_{i=1}^{\KI}}{\textstyle\frac{1}{(x^+-x_i^-)(\frac{1}{x^+}-x_i^+)}\sqrt{\frac{x^-_i}{x^+_i}}}\times\\
\nonumber
&\times & 
\frac{\displaystyle{\prod_{i=1}^{\KII}}{\textstyle \frac{(y_i-x^-)(y_i-\frac{1}{x^+})}{(y_i-x(v+\frac{i}{g}(a-2k) ))\big(y_i-\frac{1}{x(v+\frac{i}{g}(a-2k) )}\big)}\sqrt{\frac{x^+}{x^-}} }  }
{
\displaystyle{\prod_{i=1}^{\KIII}}(w_i-v-\frac{i}{g}(a-2k+1))(w_i-v-\frac{i}{g}(a-2k-1))}
\, .
\eea

\normalsize
\noindent
Here  the first and the second line correspond to the first and the second line in eq.(\ref{eqn;FullEignvalue3}), respectively. In such a representation dualization of 
$T_{a,1}^{\sl(2)}$ is straightforward, one has just to use eq.(\ref{maindual}) with proper variables $a$ and $b$.  In this way we find the expression for the transfer matrix $T_{a,1}^{\sl(2)}$
in terms of dual variables $\tilde{y}_i$, $i=1,\ldots, \KIIt$. We further change $\tilde{y}_i\to y_i$, $\KIIt\to \KII$ and denote the corresponding transfer
matrix  as $T_{a,1}^{\su(2)}$. It reads\footnote{Note that eq.(\ref{Ta1sl2}) contains a factor $\left(\frac{x^+}{x^-}\right)^{\frac{1}{2}\KII}$. When dualizing, one should express 
$\KII$ via $\KIIt$, which gives $\left(\frac{x^+}{x^-}\right)^{\KII}=\left(\frac{x^-}{x^+}\right)^{\KIIt}\left(\frac{x^+}{x^-}\right)^{\frac{1}{2}\KI+\KIII}$.}
\bea
\begin{aligned}
T_{a,1}^{\su(2)}(v)&= {\textstyle \left(\frac{x^+}{x^-}\right)^{\sfrac{\KI}{2}}  }
\displaystyle{ \prod_{i=1}^{\KII}} {\textstyle  \frac{y_i-x^+}{y_i-x^-} \sqrt{\frac{x^-}{x^+}}  }\left[
{\textstyle  P(x^-)} \prod_{i=1}^{\KI}{\textstyle{\frac{1}{x^+-x_i^-}\sqrt{\frac{x^-_i}{x^+_i}} }  }\, \displaystyle{\prod_{i=1}^{\KIII}}{\textstyle{\frac{1}
{x^-(w_i-v-\frac{i}{g}(a-1))}}}
\right.
\\
&-{\textstyle P\left(\frac{1}{x^+}\right)}
\prod_{i=1}^{\KI}{\textstyle\frac{x^--x_i^-}{(x^+-x_i^-)(\frac{1}{x^+}-x_i^+)}}
\displaystyle{\prod_{i=1}^{\KII}}{\textstyle\frac{v-\nu_i-\IG a}{v-\nu_i+\IG a}}\displaystyle{\prod_{i=1}^{\KIII}}{\textstyle \frac{x^+}{w_i-v+\frac{i}{g}(a-1)} }\\
&-{\textstyle{P(x^-)}} {\textstyle{P\left(\frac{1}{x^+}\right)}}\displaystyle{\prod_{i=1}^{\KI}}{\textstyle \frac{1}{(x^+-x_i^-)\big(\frac{1}{x^+}-x_i^+\big)}\sqrt{\frac{x^-_i}{x^+_i}}}\, \times\\
& \times
\left.
\sum_{k=1}^{a-1}\frac{\displaystyle{\prod_{i=1}^{\KII}}{\textstyle \frac{v-\nu_i+\frac{i}{g}(a-2k)}{v-\nu_i+\frac{i}{g}a}}  }
{\, \, \,  \,  \displaystyle{\prod_{i=1}^{\KIII}} (w_i-v-\frac{i}{g}(a-2k+1))(w_i-v-\frac{i}{g}(a-2k-1))\frac{x^-}{x^+}
}\right]\, .\end{aligned}
\eea
The transfer matrix above is $T_{a,1}$ in the $\su(2)$ grading. The number $\KII$ should be understood here as a number of $y$-roots which the corresponding state
exhibits in the $\su(2)$ grading. 

\smallskip

Now we turn our attention to the transfer matrix $T_{1,s}$. To obtain the $\sl(2)$ graded version of this transfer matrix, one can apply the complex conjugation
to $T_{a,1}^{\su(2)}$ regarding the roots $u,y,w$ as real and then pick up a certain normalization factor.   For real $u$ complex conjugation 
transforms $x^+_i$ into $x^-_i$ and vice versa. This motivates us to introduce a polynomial \bea P_c(y)&=&\pI
(y-x_i^+)\smp\pIII y\Big(w_i-y-\frac{1}{y}-\IG\Big)\nonumber\\
&-&\pI (y-x_i^-)\pIII y\Big(w_i-y-\frac{1}{y}+\IG\Big) \, 
\label{pol1} \eea
which is complex conjugation of $P$ defined in eq.(\ref{pol}).  Using this polynomial and replacing $a\to s$, we obtain the following expression for the transfer matrix 
$T_{1,s}^{\sl(2)}$
\bea
\begin{aligned}
T_{1,s}^{\sl(2)}(v)&= \mathscr{M}_{s}\, \displaystyle{ \prod_{i=1}^{\KII}} {\textstyle  \frac{y_i-x^-}{y_i-x^+} \sqrt{\frac{x^+}{x^-}}  }\left[
{\textstyle  P_c(x^+)} \prod_{i=1}^{\KI}{\textstyle{\frac{1}{x^--x_i^+}\sqrt{\frac{x^+_i}{x^-_i}} }  }\, \displaystyle{\prod_{i=1}^{\KIII}}{\textstyle{\frac{1}
{x^+(w_i-v+\frac{i}{g}(s-1))}}}
\right.
\\
&-{\textstyle P_c\left(\frac{1}{x^-}\right)}
\prod_{i=1}^{\KI}{\textstyle\frac{x^+-x_i^+}{(x^--x_i^+)(\frac{1}{x^-}-x_i^-)}}
\displaystyle{\prod_{i=1}^{\KII}}{\textstyle\frac{v-\nu_i+\IG s}{v-\nu_i-\IG s}}\displaystyle{\prod_{i=1}^{\KIII}}{\textstyle \frac{x^-}{w_i-v-\frac{i}{g}(s-1)} }\\
&-{\textstyle{P_c(x^+)}} {\textstyle{P_c\left(\frac{1}{x^-}\right)}}\displaystyle{\prod_{i=1}^{\KI}}{\textstyle \frac{1}{(x^--x_i^+)\big(\frac{1}{x^-}-x_i^-\big)}\sqrt{\frac{x^+_i}{x^-_i}}}\times\\
& \times
\left.
\sum_{k=1}^{s-1}\frac{\displaystyle{\prod_{i=1}^{\KII}}{\textstyle \frac{v-\nu_i-\frac{i}{g}(s-2k)}{v-\nu_i-\frac{i}{g}s}}  }
{\, \, \,  \,  \displaystyle{\prod_{i=1}^{\KIII}(w_i-v+\frac{i}{g}(s-2k+1))(w_i-v+\frac{i}{g}(s-2k-1))\frac{x^+}{x^-}}
}\right]\, .\end{aligned}
\eea
Here the overall normalization factor  \bea \nonumber
\mathscr{M}_{s}={\textstyle{(-1)^s }}\pI {\textstyle{\left(\frac{x^-_i}{x^+_i}\right)^{\sfrac{s}{2}}\frac{x^--x^+_i}{x^+-x^-_i}   }}
\prod_{k=1}^{s-1}
{\textstyle{\frac{x(v+\frac{i}{g}(s-2k))-x^+_i}{x(v-\frac{i}{g}(s-2k))-x^-_i}}}\, \eea 
has been found by requiring that $T_{1,s}^{\sl(2)}$ reproduces $T_{a,1}^{\sl(2)}$ through Bazhanov-Reshetikhin formula. Note that $\mathscr{M}_{s}$ is also
a gauge transformation. 

\smallskip 

Finally, we present for completeness the transfer matrix $T_{1,s}$ in the $\su(2)$ grading. It is obtained by complex conjugation of $T_{a,1}^{\sl(2)}$ and picking up 
a proper normalization factor :
\bea
\nonumber
T_{1,s}^{\su(2)}(v)&=&\mathscr{M}_s{\textstyle \left(\frac{x^+}{x^-}\right)^{\frac{\KI}{2}}}\displaystyle{\prod_{i=1}^{\KII} \textstyle{\frac{y_i-x^+}{y_i-x^-}\sqrt{\frac{x^-}{x^+}}}}\left[{\textstyle{P_c(x^-)}}\,
\prod_{i=1}^{\KI}{\textstyle\frac{ 1}{x^--x_i^+}\sqrt{\frac{x^+_i}{x^-_i}}}
\,  \displaystyle{\prod_{i=1}^{\KIII}}{\textstyle\frac{1}{x^-(w_i-v+\frac{i}{g}(s-1))}}  \right.\\
\label{T1ssu2}
&-&{\textstyle{P_c\left(\frac{1}{x^+}\right)}}
\displaystyle{\prod_{i=1}^{\KII}{\textstyle\frac{v-\nu_i-\frac{i}{g}s}{v-\nu_i+\frac{i}{g}s}}
\prod_{i=1}^{\KI}{\textstyle \frac{x^+-x_i^+}{(x^--x_i^+)(\frac{1}{x^-}-x_i^-)}}
}\displaystyle{\prod_{i=1}^{\KIII}{\textstyle\frac{x^+}{w_i-v-\frac{i}{g}(s-1)}}}\\
\nonumber
&-&\sum_{k=1}^{s-1}{\textstyle{P_c\Big(\frac{1}{x(v-\frac{i}{g}(s-2k))}\Big)P_c\Big(x(v-\frac{i}{g}(s-2k))\Big)}}
\displaystyle{\prod_{i=1}^{\KI}}{\textstyle\frac{1}{(x^--x_i^+)(\frac{1}{x^-}-x_i^-)}\sqrt{\frac{x^+_i}{x^-_i}}}\times\\
\nonumber
&\times & \left.
\frac{\displaystyle{\prod_{i=1}^{\KII}}{\textstyle \frac{v-\nu_i-\frac{i}{g}s}{v-\nu_i-\frac{i}{g}(s-2k) }}  }
{
\displaystyle{\prod_{i=1}^{\KIII}}(w_i-v+\frac{i}{g}(s-2k+1))(w_i-v+\frac{i}{g}(s-2k-1))}\right]
\, .
\eea
This completes our discussion of the duality transformation for transfer matrices.


\end{document}